\documentclass[a4paper,11pt]{article}
\usepackage{pos}

\newcommand{\appropto}{\mathrel{\vcenter{
  \offinterlineskip\halign{\hfil$##$\cr
    \propto\cr\noalign{\kern2pt}\sim\cr\noalign{\kern-2pt}}}}}

\title{Anisotropy of Cosmic Rays and Chaotic Trajectories in the Heliosphere}
 \ShortTitle{Anisotropy of Cosmic Rays and Chaotic Trajectories in the Heliosphere}

\author*[a,b,c]{V. L\'opez-Barquero}
\author[b,c]{P. Desiati}

\affiliation[a]{University of Cambridge, UK.}
\affiliation[b]{University of Wisconsin - Madison, U.S.A.}
\affiliation[c]{Wisconsin IceCube Particle Astrophysics Center, U.S.A.}

\emailAdd{vlbarquero@icecube.wisc.edu}
\emailAdd{desiati@icecube.wisc.edu}

\abstract{As cosmic rays (CRs) propagate in the Galaxy, they can be affected by magnetic structures that temporarily trap them and cause their trajectories to display chaotic behavior, therefore modifying the simple diffusion scenario. When CRs arrive at the Earth, they do so anisotropically. These chaotic effects can be a fundamental contributor to this anisotropy. Accordingly, this requires a comprehensive description of chaos in trapping conditions since it is necessary to assess their repercussions on the CR arrival directions. This study utilizes a new method described in López-Barquero and Desiati (2021) to characterize chaotic trajectories in bound systems. This method is based on the Finite-Time Lyapunov Exponent (FTLE), a quantity that determines the levels of chaos based on the trajectories' divergence rate. The FTLE is useful since it adapts to trapping conditions in magnetic structures or even propagating media changes. Here, we explore the effects that chaos and trapping can have on the TeV CR anisotropy. Concretely, we apply this method to study the behavior of CRs entering the heliosphere. Specifically, how the distinct heliospheric structures and CR impinging directions from the ISM can affect chaos levels. The heliosphere has an intrinsic directionality that affects CRs differently depending on where they enter it. This feature causes preferential directions from which particles tend to be more chaotic than others. This eventually translates into changes in the arrival maps which are not uniformly distributed. Instead, we expect sectors in the map to change separately from others, creating a time variation that could be detected. Consequently, this result points to the idea that time-variability in the maps is essential to understanding the CR anisotropy's overall processes.  
}

\FullConference{37$^{\rm{th}}$ International Cosmic Ray Conference (ICRC 2021)\\
		July 12th -- 23rd, 2021\\
		Online -- Berlin, Germany}

\begin{document}
\maketitle

\section{Introduction}

Cosmic rays of Galactic origin are detected on the Earth with anisotropy in their arrival directions. This anisotropy ($10^{-3}$ in relative intensity) has been measured by multiple experiments~\citep{Abeysekara_2019,aartsen_2013chaos}; nonetheless, a complete explanation for it is still eluding us. This work will explore the contributions that chaotic trajectories of trapped cosmic rays can have on it.  Specifically, how coherent structures, e.g., the heliosphere, can significantly impact how particles propagate and ultimately the directions that they are detected.  

\section{Chaotic Trajectories and Coherent Magnetic Structures}

In order to assess the chaotic effects on particles' trajectories and how they are affected by being temporarily trapped in coherent structures, we develop a new method for characterizing chaos and construct a toy model that will replicate the trapping conditions in these magnetic structures. 

This model is based on the Finite-Time Lyapunov exponent (FTLE):

\begin{equation}\label{FTLE}
\lambda (t,\Delta t)=   \frac{1}{\Delta t}   \ln \left [ \frac{d(t+\Delta t)}{d(t) } \right ],
\end{equation}
where $\Delta t$ is the time interval for the calculation and d(t) is the the distance between two particles at time t. Thus, the FTLE can measure the level of chaos based on the trajectories’ divergence rate. The usefulness of this quantity is that it can adapt to the temporarily trapped conditions that can emerge due to the interaction with coherent magnetic structures. 

In order to reproduce the trapping conditions in a magnetic field, we created a model that consists of an axial-symmetric magnetic bottle with magnetic time-perturbations added: 

\begin{equation}
B_{y} = \frac{\Delta B}{B}\,\sin(k_p x-\omega_p t)\,e^{-\frac{1}{2}\left(\frac{z}{\sigma_p}\right)^2},
\label{eq:pert}
\end{equation}
where $k_p = \frac{2\pi}{L_p}$ and $\omega = \frac{2\pi v_p}{L_p}$. 

The specific characteristics of the model used in this work are based on heliospheric conditions. The magnetic bottle is based on the mirroring effect that particles experience as they bounce out of the flanks of the heliosphere.  The time perturbations replicate the effects of magnetic field reversals induced by the 11-year solar cycles.

\section{Discussion}

Once we propagate particles in this system. we found a correlation between the Finite-time Lyapunov exponent (FTLE), i.e., the chaotic behavior of the particles, and the escape time from the system. This correlation is given by a power law:

\begin{equation}
\lambda_{\tiny{FTLE}} = \beta\, t_{esc}^{-1.04 \pm 0.03} .
\label{eq:powerlaw}
\end{equation}

\begin{figure}
	\centering
	\includegraphics[width=1\linewidth]{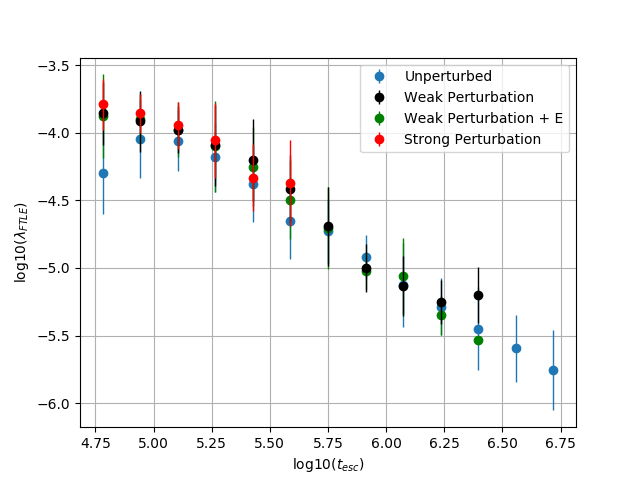}
	\caption[four cases]{  \textit{The Finite-Time Lyapunov exponent, $\lambda_{\tiny{FTLE}}$, vs. the escape time from the system, $t_{esc}$, for four different systems}. The blue points represent the unperturbed system. The black points correspond to the profile of the weak-perturbation system. The green ones show the weak perturbation plus electric field, and the red ones the strong-perturbation system. Note that the power-law remains the same even when perturbations are introduced.}
	\label{fig:finitelyapunovvsescapetimefourcaseschi2500}
\end{figure}

One remarkable feature found for these systems is that the same power-law persists even if perturbations are introduced in the systems, see figure \ref{fig:finitelyapunovvsescapetimefourcaseschi2500}.

To derive information that will help us elucidate the observations, the Finite-Time Lyapunov exponents and escape times are plotted in arrival distribution maps. In these maps, regions with different chaotic behavior emerge, which can have an impact on the observations. For example, this can be a source of time-variability in the anisotropy maps.

%
%
%

\end{document}